# A report of (topological) Hall anomaly two decades ago in $Gd_2PdSi_3$, and its relevance to the history of the field of Topological Hall Effect due to magnetic skyrmions[#]

E.V. Sampathkumaran*

*Tata Institute of Fundamental Research, Homi Bhabha Road, Colaba, Mumbai 400005, India*

Abstract

The area of 'magnetic skyrmions' – with a potential for revolution in certain applications - in condensed matter physics is considered about a decade old. In this article, we draw the attention of the community to the recent work of Kurumaji et al [(Science 365, 914(2019)] establishing magnetic skyrmion lattice behavior in $Gd_2PdSi_3$. We consider it important to bring out the two decades old history, in particular Hall transport anomalies characterizing magnetic skyrmions.

Following rich traditions of our exhaustive investigations on rare-earth intermetallics, making original contributions to many phenomena over a period of several decades, detailed studies on $AlB_2$-derived ternary (2-1-3) rare-earth intermetallics with triangular arrangement of magnetic ions were pioneered by us more than two decades ago. At that time, we were prompted to investigate the title compound, triggered by the observation of abnormal transport behavior of many materials (in different families) containing Gd and other 'normal' rare-earth ions (like Tb, Dy etc) – which are traditionally considered uninteresting.

The key experimental data in support of the skyrmion lattice state that Kurumaji et al present is the observation of 'topological Hall resistivity' in an intervening field range, following two metamagnetic transitions. In addition, the values are 'giant' compared to those known for other magnetic skyrmions. We point out here that such features in Hall resistivity data on the same compound, arising from Gd 4f magnetism, can be found in our publication in 1999 (along with the metamagnetic transitions) and we expressed difficulties in explaining the results at that time. In view of above, we propose to the scientific community that the compound $Gd_2PdSi_3$, ex post facto, can be considered as the first experimental demonstration for 'giant topological Hall effect' in this field, discovered two decades ago.

We take this opportunity to share some thoughts, offering some clues for further investigations, in particular to identify novel magnetic skyrmions and to find the same to enable applications.

*E-mail: sampathev@gmail.com

#A brief e-letter can be found at the Science journal site:
https://science.sciencemag.org/content/365/6456/914/tab-e-letters

Following the experimental discovery of 'vortex'-like nanometric spin textures, called 'magnetic skyrmions', in which the spins are oriented in all directions wrapping up a sphere, in MnSi [1] and $Fe_{0.5}Co_{0.5}Si$ [2], around 2009, this area of research has been attracting considerable attention in condensed matter research due to potential spintronic applications as well as for next generation information storage due to very low current density (that is, ultra low threshold) for current-driven motion. This concept was theoretically advanced some years earlier [see, for instance, the recent reviews, Refs. 3 and 4]. While small angle neutron scattering [1] and Lorenz transmission microscopy [2] have been the primary tools at the beginning of this field to observe spin textures, it was soon realized [see, for instance, Refs. 5 and 6] that the key characteristic of magnetic skyrmions is that there is an additional contribution to Hall resistivity ($R_H$) in the magnetic field and temperature range in which such a phase exists (over and above well-known normal and anomalous contributions in magnetic materials). Such an unconventional extra contribution is called 'Topological Hall Effect (THE)'. Thus, the observation of THE in an intermediate magnetic field emerged as a hallmark of the magnetic skyrmion behavior. How such an unconventional Hall contribution arises has been explained, for instance, in a recent review article by Kim [4], in terms of the emergent magnetic field due to the strong coupling between the spins of skyrmions and conduction electrons; since the local magnetization varies at each point, the conduction electrons continuously feel the force which attempts to reorient. It should however be noted that the values of the extra Hall resistivity reported till recently fall in the nΩcm range in any previously known magnetic skyrmion. Another belief in the literature is that such topologically nontrivial spin textures should be observed among non-centrosymmetric materials, as by now well-known Dzyaloshinskii-Moriya (DM) interaction in such systems favors twisting of spins arrangement. However, in recent years, there are a few theories [6-8] predicting such magnetic textures in the event of geometrical frustration of magnetic moments. Very recent reports of observation of magnetic skyrmions in the centrosymmetric compounds with magnetic triangles, $Gd_2PdSi_3$ [9] and $Mn_3Sn$ [10], are considered as experimental proof for those theories. In short, the field of 'magnetic skyrmions' is considered to be about a decade old from the angle of experimental results.

*In this article, we want to recall the history of $Gd_2PdSi_3$ in the context of the recent claim of (giant) THE by Kurumaji et al [9] on $Gd_2PdSi_3$, attributed to skyrmion lattice behavior in this centrosymmetric structure (endorsed by resonant x-ray scattering studies). The values of Hall resistivity reported by these authors for the skyrmion phase are quite huge in the µΩcm range, unlike that known for other magnetic skyrmions. We would like to point out that we have reported exactly the same Hall features in 1999 for this compound (see Fig. 1), attributed to dominant Gd4f magnetism.* We consider it important for the benefit of the community to recall the two decades old history of this compound [11-13], making additional comments for future work.

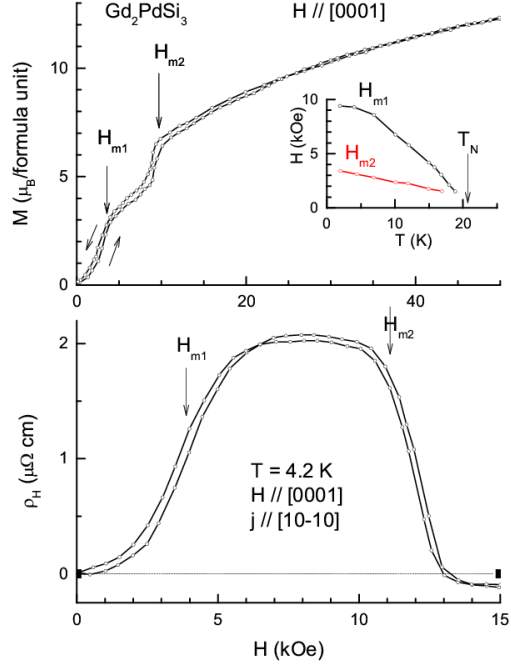

Fig. 1: Isothermal magnetization (at 2 K) and Hall resistivity (at 4.2 K) as a function of magnetic field for H//[0001] for $Gd_2PdSi_3$ (data extracted from our publication in 1999, Ref. 13). The observation being stressed – relevant to the main point of this article – is that the (giant) Hall resistance for intervening fields across metamagnetic transitions resembling that reported by Kurumaji et al recently [9] – can be seen in the bottom figure. Metamagnetic transition fields $H_{m1}$ and $H_{m2}$ reported by us are also marked and their temperature dependencies, shown in Ref. 13, are also plotted in the inset.

We initially investigated the polycrystalline form of this compound more than two decades ago, and already found Hall anomalies, measured with 10kOe, changing its sign (to positive) well below Néel temperature ($T_N$= 21 K) unlike in $Lu_2PdSi_3$, as shown in figure 2, reproduced from Ref. 11. This is apart from other features [12] like the existence of antiferromagnetic and ferromagnetic correlations without consistent spin-glass anomalies already indicating complexity of magnetism, another subtle magnetic anomaly (as inferred from $^{155}$Gd Mössbauer spectra), negative temperature coefficient of electrical resistivity ($\rho$) and huge magnetoresistance not only in the magnetically ordered state but also well above $T_N$ (see figure 3, data reproduced from Ref. 12). These interesting features, in particular in Hall data, immediately prompted us to pursue the single crystal investigations (thanks to the collaboration offered by the group of H. Sato, Tokyo Metropolitan University) for a better understanding of the features. The key point we would like to stress, that is transparent in the figure 1 (bottom, data reproduced from Ref. 13) to the reader, is that Hall resistivity shoots up at the first metamagnetic transition and falls at the next transition; in other words, it goes without saying explicitly that it happens for an intervening field range (at 4.2 K, H ~ 3 – 12 kOe) across two metamagnetic transitions for the geometry H//c. Thus, "distinct anomalies" at the two metamagnetic transition fields was brought out. *We had argued that it was not possible to explain the above-mentioned Hall resistivity behavior as a function of H by the conventional wisdom in the literature at that time* (that is, an explanation in terms of the sum of ordinary Hall effect and conventional anomalous magnetic scattering contributions). These features in single crystals are apart from the observations of anisotropic electrical resistivity, magnetoresistance and magnetization anomalies, another subtle magnetic feature near 15 K (supporting Mössbauer spectral features stated above), as well as Hall anomalies even in the paramagnetic state near 100 K (see Fig. 5 in Ref. 13), thereby stressing in our publications on the richness and anomalous magnetism of this compound.

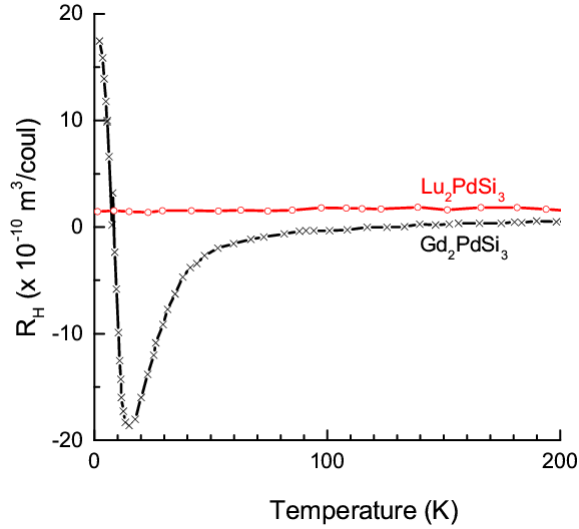

Fig. 2: The temperature dependence of Hall coefficient of polycrystalline $Gd_2PdSi_3$ and $Lu_2PdSi_3$, from the data extracted from our work in 1999 Ref. 11. It is clear that there is a sign reversal of Hall anomaly around 10 K in the magnetically ordered state ($T_N$= 21 K), when measured in a field of 10 kOe, whereas in Lu analogue, there is no worthwhile feature. This comparison revealed the dominant role of Gd 4f magnetism on Hall behavior.

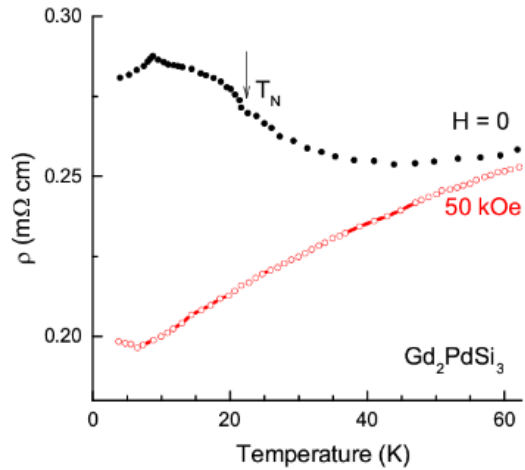

Fig. 3: Electrical resistivity as a function of temperature for polycrystalline $Gd_2PdSi_3$ in zero field and in 50 kOe, extracted from our publication in 1998 (Ref. 12). The $d\rho/dT$ is negative below $T_N$ (without showing the drop due to the loss of spin-disorder contribution), indicating complex magnetic phase. It is to be noted that there is a minimum in $\rho(T)$ well above $T_N$, uncharacteristic of "normal" rare-earths like Gd, which gets suppressed by the application of magnetic field. We here raise the question whether this is due to skyrmions building gradually with decreasing temperature even in zero field (just as pseudogap formation in high temperature superconductors), contributing to electronic scattering; if this is true, such a scattering contribution appears to suppressed with the applied field.

We do not want to dwell on a possible scenario indicated at that time (in the absence of the currently recognized concept of 'skyrmions' in magnetic materials). Nevertheless, we want to share some thoughts with respect to that speculation. By comparison of Hall behavior of Lu and Gd compounds, we had clearly stressed [11, 13] on the dominant 4f magnetic contribution for the observed Hall behavior for the Gd case. In addition, Fermi surface effects (like changes in magnetic superzone gaps, see, for instance, Ref. 14) across metamagnetic transitions were indicated [11, 13], which is expected for complicated magnetic structures. The conduction electrons are strongly coupled to skyrmions as stated in the introductory paragraph, as these are made up of large moment carrying Gd ions, and the emergent field can influence the magnetic Brillouin-zone gaps. Therefore, in our opinion, in metallic skyrmions, the contributions due to local moment and Fermi surface topology to Hall resistivity are coupled and can not be ignored. Any change in magnetic lattice periodicity, which can happen by a variation of temperature or magnetic field, is bound to influence Fermi surface topology. It may also be remarked that subsequent Angle Resolved Photoemission Spectroscopic studies following this speculation indeed established decisive role of the Fermi surface geometry on the complex magnetism of this compound [15]; therefore one would expect that the Fermi surface topology changes for the magnetic phase in the intermediate field are also expected. We prompt future investigations to throw some light on whether this factor also matters for the observed Hall anomaly.

From above, it is abundantly clear that the Hall resistance behavior in particular, the jumps (to a giant value) in an intervening field range in the magnetically ordered state, was crucial (also see the Note by the Editor, Science, Ref. 16) to the conclusions of Kurumaji et al, and that we reported the same Hall features two decades ago for the same compound (viz., in $Gd_2PdSi_3$) [Ref. 11, 13]. In a private communication, Kurumaji kindly acknowledged that our work is the first one reporting such a Hall anomaly. Kurumaji et al [9] of course mention in the introduction of their first publication that their work was motivated by 'nonmonotonic variations of longitudinal and transverse transport properties', reported by us. In our opinion, it is difficult for the readers to infer that this statement refers to the above-mentioned Hall anomaly and to trace the history of this feature to the 1998-1999 period work. More so because, the subsequent paper [17], posted in arXiv, does not take note of the old literature on this compound. Therefore, for the benefit of the readers, we feel the need to bring out that, to our knowledge, the discovery of what is now called 'Topological Hall Effect' (in the context of magnetic skyrmions) in the literature, is much older than what is believed currently. *In view of this, we appeal to the community that this compound be considered, ex post facto, as the first example for 'topological Hall effect' behavior, arising from 'magnetic skyrmion'.*

***Finally, we would like to add some general remarks, also making some suggestions for future work.***

This 'rare-earth (R) transition-metal (TM) silicides/germanides' ternary family is derived from $AlB_2$-derived hexagonal structure [18, 19]. This structure is of honey-comb type, in which TM-Si(Ge) layers are sandwitched by R layers. R ions form triangular lattice, thereby favoring geometrically frustrated magnetism in the event of intersite antiferromagnetic interaction. The Pd-based family was synthesized by Kotsanidis et al [20] reporting preliminary magnetization results. We initiated detailed studies on these ternary families immediately after that report, as the ordered replacement of boron site by Pd and Si can result in the formation of superstructures with consequences on physical properties. As a consequence of this chemical order, two crystallographic sites for R are generated, as demonstrated originally from a microscopic tool ($^{151}$Eu Mössbauer effect) on $Eu_2PdSi_3$ [21] and by neutron diffraction [22]. We initially established Kondo lattice anomalies in $Ce_2RhSi_3$ and $Ce_2PdSi_3$ [23], as a continuation of our efforts at TIFR over a period of some decades on Ce, Eu and Yb alloys to probe 4f localization-delocalization phenomena, like Kondo lattice anomalies due to a competition between the well-known RKKY indirect exchange interaction and the Kondo effect, valence fluctuations etc. In early 1990s, we made unique contributions in many other families with different crystal structures, by revealing exotic thermal and transport anomalies even on materials containing other 'normal' rare-earths like Gd. Tb, Dy etc. This prompted us to investigate such 'normal' rare-earth compounds in this ternary family as well. The point to be stressed here is that, we found a variety of exotic properties, not only in Ce based-systems, but also on such 'normal' rare-earth systems, raising many interesting questions with respect to magnetic and transport properties of solids (including superconductivity in Y-based systems, geometrically frustrated magnetism due to triangular arrangement of rare-earth ions, magnetocaloric effect [24], large magnetoresistance even in the paramagnetic state [25]). Interested readers may see recent references [15, 26] and articles cited therein for a partial list. Though a few groups (see, for instance, Refs. 22, 27, 28) published some work on this ternary family subsequent to our work, a deeper understanding of the properties of this 2-1-3 family remains largely unexplored, and it is worthwhile to pursue a systematic study of the same in the context of THE/magnetic-skyrmions, which possibly could throw more light on this phenomenon. We continue our investigations keeping this in mind.

At this juncture, it is important to mention that $Gd_2PdSi_3$ was shown to show large magnetocaloric effect in the vicinity of $T_N$, and there are sign reversals in isothermal magnetic entropy change for a field of 5 or 10 kOe (which falls in the "intervening" field range according to the terminology in the current literature of magnetic skyrmions) as a function of temperature [24], already supporting the idea of competing antiferromagnetic and ferromagnetic couplings, leading to canting of magnetic moments.

In particular, we had emphasized that the heavy rare-earth members provide an opportunity to study exotic phenomena without much complications arising from the itinerant magnetism and the Kondo effect. It was noted that the triangular arrangement of antiferromagnetically coupled rare-earth ions results in features attributable to geometrical frustration of magnetism, not only in the magnetically ordered state, but also in the temperature range prior to long range magnetic order in this hexagonal structure. For instance, $Gd_2PdSi_3$ is characterized [12] by a Kondo-like minimum above $T_N$ (see Fig. 3), unexpected for rare-earths containing well-localized 4f electrons. We proposed that a novel magnetic precursor effect, possibly electron localization, the root-cause of which lies in magnetism, even in metallic systems may have to be thought of to explain upturns (or an extra contribution adding to lattice part) in electrical resistivity before long range magnetic order sets in. However, this feature remained a puzzle for nearly two decades. In fact, subsequently, we found such upturns / extra contributions in many 'normal' heavy rare-earth systems including Gd and the list of such anomalous systems is endless. Recently, Wang et al [29] proposed a RKKY-based model to explain such low-temperature upturns as a result of magnetic frustration. Now that the magnetic skyrmions are found in this system, which is also due to geometrical frustration, we raise a question whether 'skyrmions' can also be invoked, possibly in the light of the theory of Ref. 29. That is, these skyrmions build gradually with a lowering of temperature even in zero field as a result of magnetic frustration, disbursing randomly in the paramagnetic matrix before a tendency to form (skyrmion) lattice (after the onset of long-range magnetic order) by applying a small field – just as pseudo-gap formation in high temperature superconductors – contributing to electrical resistivity. If so, we pose the question whether the electronic scattering contribution from such "dilute" or "random skyrmions" (also possibly due to gradually varying sizes) increases with decreasing temperature, just as the resistivity minimum was originally reported in dilute magnetic impurity (Kondo) systems. It is of interest to address this issue ("single skyrmion behavior to skyrmion lattice behavior" similar to "Kondo impurity behavior to Kondo lattice behavior") theoretically as well. We do not know whether observable magnetoresistance [see, Fig. 2 in Ref. 11] and Hall constant anomalies [see, Fig. 5, Ref. 13], occurring in the range 60 - 100 K even in the paramagnetic state signal the beginning of 'liquid' skyrmion formation. If this conjecture is correct, the question remains to be answered why such independent skyrmions do not form the lattice in zero field in the magnetically ordered state. In this respect, one can even study dilution effects, e.g., $Gd_{2-x}Y_xPdSi_3$, to understand this transformation [12, 30].

We reiterate that the magnetic precursor effects in bulk properties were reported earlier in many heavy rare-earth members of other families as well [31], which primarily motivated us to investigate $Gd_2PdSi_3$. There is a voluminous background work reported by us on such traditionally 'uninteresting' rare-earth ions, which motivated us to study this compound in detail. Subsequently, we found many other similar heavy rare-earth members in different families after that work. Based on our magnetic, electrical and magnetotransport anomalies, we believe that there are many other families (where we found low-field metamagnetism and multiple metamagnetic transition with exotic hysteretic effects) worth pursuing for THE and skyrmion physics. Some of the heavy rare-earths order magnetically close to room temperature, and if THE is found in these, could be potential candidates for applications. Usually, in the rare-earth literature, the research community does not usually probe such 'normal' rare-earths in depth, as the 4f electrons, which are core-like in these were not expected to lead to any conceptually new phenomena, though there was some interest in their compounds in the fields of magnetocaloric effect, permanent magnets etc.

To justify the above statement, we make the following observation. We first reported [32] magnetic and magnetoresistance anomalies in a metallic kagome lattice, $Gd_3Ru_4Al_{12}$, a few years ago, suggesting competition from various collective magnetic interactions due to geometrical frustration, low-field metamagnetic transitions and a magnetic feature below 10 K (in addition to $T_N$= 18.5 K). It is satisfying to note that this Gd compound as well motivated Tokura's group to search [33] for skyrmion lattice behavior with a giant THE successfully. In light of such reports, triggered by our own original results, we believe that such 'rare-earth intermetallics' could be a very good playground for skyrmion physics and giant THE.

In particular, a large number of derivatives are possible by varying transition metal ions in the 2-1-3 family, and thus the opportunities provided by this ternary family are even more than those known for (layered) $ThCr_2Si_2$ tetragonal structure - one of the most commonly studied rare-earth families in which several exotic phenomena were discovered in the literature during last five decades. Such exhaustive studies (including solid solutions) are still underway.

It is also of interest to find suitable systems, where the field/temperature-induced skyrmion is arrested under ambient conditions after travelling through this skyrmion region. Such an arrest of ferromagnetic or antiferromagnetic phases upon supercooling/superheating across first-order transitions and consequent thermomagnetic history effects had been proposed and demonstrated in the past literature on many materials by the group of Chaddah and Roy [34]. This could be a route to attain materials for applications at room temperature. In short, we urge the community to carry out in depth investigations on such "normal" rare-earth compounds as well from the angle of 'magnetic skyrmions', taking clues from above points.

It is not clear whether the oscillatory nature of RKKY interaction in rare-earth metallic systems coupled with strong coupling under favorable circumstances plays a key role in reorientation of the spin direction to result in the formation of skyrmions [35].

**Acknowledgements:**

I would like thank Takashi Kurumaji for his personal communication kindly endorsing that the nonmonotonic (giant) Hall resistivity behavior reported by us on this compound two decades ago is the first one. I thank the then graduate students Indranil Das and Roop Mallik during initial stages of the work on 2-1-3 intermetallics as well as to study magnetic and transport anomalies of other 'normal' rare-earths. Subsequently several students and post-doctoral fellows were involved, and P.L. Paulose, Kartik K Iyer, K. Maiti, and many others from abroad (e.g., the groups of C. Laubschat and G. Wortmann in Germany) collaborated with me from time to time over a period of 25 years to study such families of compounds with different experimental methods. Their contributions are gratefully acknowledged. Spontaneous agreement to collaborate on single crystals by the co-authors at Tokyo Metropolitan University, and in particular, S.R. Saha (the then graduate student, presently at the University of Maryland) for his enthusiastic interaction with me during single crystal studies on $R_2PdSi_3$ (R= Ce, Gd) in late 1990s are acknowledged. I am thankful to the collaborators at the single crystal growth group at IFW, Dresden, for pursuing single crystal investigations on many other members of this family till now.

References:

1. S. Mühlbauer et al., Science 323, 915 (2009).
2. X.Z. Yu et al., Nature 465, 901 (2010).
3. C.D. Batista et al., Rep. Prog. Phys. 79, 084504 (2016).
4. Bom Soo Kim, J. Phys.: Condens. Matter 31, 383001 (2019).
5. A. Neubauer et al., Phys. Rec. Lett. 102, 186602 (2009).
6. Y. Li et al., Phys. Rev. Lett. 110, 117202 (2013); M. Lee et al., Phys. Rev. Lett. 102, 186601 (2009).
7. A.O. Leonov and M. Mostovoy, Nat. Commun. 6, 8275 (2015).
8. T. Okubo, S. Chung, and H. Kawamura, Phys. Rev. Lett. 108, 017206 (2012).
9. T. Kurumaji et al., Science 365, 914 (2019).
10. P.K. Rout et al., Phys. Rev. B 99, 094430 (2019).
11. R. Mallik et al., Pramana – J. Phys. 51, 505 (1999).
12. R. Mallik et al., Europhys Lett. 41, 315 (1998).
13. S.R. Saha et al., Phys. Rev. B60, 12162 (1999).
14. I. Das and E.V. Sampathkumaran, Phys. Rev. B 49, 3972 (1994); I. Das et al., Phys. Rev. B 44, 159 (1991).
15. D.S. Inosov et al., Phys. Rev. Lett. 102, 046401 (2009).
16. See, the note by the editor, J. Stajic, Science, 365, 880 (2019).
17. Max Hirschberger et al, arXiv:1910.06027v1.
18. B. Chevalier et al, J. Alloys and Compd. 233, 150 (1996).
19. R.E. Gladyshevskii, K. Censual, and E. Parthe, J. Alloys and Compd. 189, 221 (1992).
20. P. A. Kotsanidis, J.K. Yakinthos, and E. Gamari-seale, J. Magn. Magn. Mater., 87, 199 (1990).
21. R. Mallik et al., 185, L135 (1998).
22. F. Tang et al., Phys. Rev. B 84, 104105 (2011).
23. I. Das and E.V. Sampathkumaran, J. Magn. Magn. Mater. 137, L239 (1994); R. Mallik and E.V. Sampathkumaran, J. Magn. Magn. Mater. 164, L13 (1996); for our subsequent single crystal work on $Ce_2PdSi_3$, see, S.R. Saha et al., Phys. Rev. B 62 (2000) 425.
24. E.V. Sampathkumaran, I. Das, R. Rawat, and Subham Majumdar, App. Phys. Lett. 77, 418 (2000).

25. R. Mallik, E.V. Sampathkumaran, and P.L. Paulose, Solid State Commun. 106, 169 (1998); Subham Majumdar, E.V. Sampathkumaran, P.L. Paulose, H. Bitterlich, W. Loser, and G. Behr, Phys. Rev. B 62, 14207 (2000);

26. K. Maiti et al., arXiv:1909.1001 and articles cited therein; M. Smidman et al, Phys. Rev. B 100, 134423 (2019).

27. M. Frontzek et al, J. Phys.: Condens. Matter 19, 145276 (2007).

28. D.X. Li, S. Nimori, Y. Shiokawa, Y. Haga, E. Yamamoto, and Y. Onuki, Phys. Rev. B 68, 012413 (2003).

29. Z. Wang et al., Phys. Rev. Lett. 117, 206601 (2016).

30. Subham Majumdar et al., J. Phys.: Cond. Matter. 11 (1999) L329.

31. R. Mallik, E.V. Sampathkumaran, P.L. Paulose, and V. Nagarajan, Phys. Rev. B 55, R8650 (1997); R. Mallik and E.V. Sampathkumaran, Phys. Rev. B 58, 9178 (1998) and articles cited therein.

32. Venkatesh Chandragiri, Kartik K Iyer, and E.V. Sampathkumaran, Intermetallics, 28, 286002 (2016).

33. Max Hirschberger et al, arXiv:1812.02553.

34. M. Manekar et al., J.Phys.: Matter 14, 4477 (2002). See, for a review, S.B. Roy and P. Chaddh, Phys. Status Solidi B 251, 2010 (2014).

35. Z. Wang and C.D. Batista informed this author after the first version was posted that they have addressed the question whether RKKY interaction stabilizes skyrmionic crystal; the result is affirmative for a question raised in this article. They plan to post a preprint soon.